# Combinatorial sputter synthesis of single-phase La(XYZ)O$_{3\pm\delta}$ perovskite thin film libraries: a new platform for materials discovery

Tobias H. Piotrowiak[1], Rico Zehl[1], Ellen Suhr[1], Benedikt Kohnen[1], Lars Banko[1], Alfred Ludwig*,[1]

Underlined authors have contributed equally.

[1]Chair for Materials Discovery and Interfaces, Institute for Materials,
Ruhr University Bochum, Universitätsstraße 150, 44780 Bochum, Germany

## Abstract

Compositionally complex perovskites provide the opportunity to develop stable and active catalysts for electrochemical applications. The challenge lies in the identification of single-phase perovskites with optimized composition for high electrical conductivity. Leveraging a recently discovered effect of self-organized thin film growth during reactive sputtering, La-Co-Mn-O and La-Co-Mn-Fe-O perovskite (ABO$_3$) thin film materials libraries are synthesized. These show phase-pure La-perovskites over a wide range of chemical composition variation for the B-site elements for deposition temperatures ≥ 300 °C. It is demonstrated that this approach enables the discovery and tailoring of chemical compositions for desired optical bandgap and electrical conductivity, and thereby opens the path for the targeted development of e.g. new high-performance electrocatalysts.

## 1 Introduction

Perovskites are an important subject of current materials research. Especially La-based perovskites are promising materials for solid oxide fuel cells[1,2] and for electrocatalytic water splitting[3,4]. However, it is a huge challenge to determine the best performing chemical compositions for the respective application, given the enormous compositional space offered by perovskites consisting of multiple elements. The general formula for perovskite is ABO$_3$:

La- and other lanthanides occupy the A-sites. The B-sites can only be occupied by transition metals such as Co, Mn, Fe, Ni, and V. The respective A-site and B-site elements can substitute each other but cannot switch their sites.[5] Elemental combinations up to high-entropy perovskites[6,7] were reported to be performant for electrocatalytic water splitting. The increase in chemical complexity provides more opportunities to optimize the materials for target applications. However, this comes at the price of a dramatically increased search space for new materials given by a quasi-infinite number of possible chemical and proportional compositions, that need to be screened for their properties. Therefore, the conventional approach, by means of subsequent fabrication of single samples with discrete composition is not efficient nor target-oriented. A more suitable approach for the screening of large material compositional spaces is combinatorial materials science. Thereby, hundreds or thousands of samples can be fabricated under identical conditions in a single process on a single substrate, a so-called materials library. These samples are then investigated for their chemical composition and other materials properties.[8,9] Suitable synthesis methods are for example inkjet printing[10,11] or gas-phase deposition processes like sputter deposition[12–14].

Sputter deposition is well-established in combinatorial materials science and has advantages over other processes such as high purity of synthesized materials and superior processing speed. Moreover, it ensures good reproducibility. Nonetheless, sputtering of poly-elemental single-phase materials libraries with compositional gradients usually works only within the ranges of the solubility limits of the respective elements in the phase and other parameters relevant for the phase formation like temperature and stoichiometry. If there is a mismatch, multiphase regions can form according to thermodynamically favored states of the lowest energy that can be achieved under these conditions. This commonly hinders the fabrication and subsequent investigation of single-phase materials in co-sputtered materials libraries with compositional gradients.

However, a single-phase formation behavior was recently observed and described by the authors for reactive hot deposition of (La-based) perovskites[15]. This phase formation behavior can be leveraged to synthesize phase-pure perovskite materials libraries for a wide range of chemical compositions of the B-sites. It can be described as a self-organized growth process based on the perovskite stoichiometry, which suppresses formation of multiphase regions in favor of single-phase perovskite.

In the present work not only the fabrication of La-Co-Mn-O and La-Co-Mn-Fe-O libraries containing expanded perovskite regions with compositional gradients regarding the B-sites will be demonstrated, but also the subsequent high-throughput characterization of the electrical resistivity and optical bandgaps of these perovskites. Taking advantage of this synthesis route and of high-throughput materials characterization techniques, discovery of novel perovskite materials for different applications can be accelerated.

## 2 Results and Discussion

### 2.1 Chemical composition and phase analysis

For reactive sputtering on substrates heated to temperatures ≥ 300 °C, crystalline single-phase perovskite thin films in the systems La-Co-Mn-O and La-Co-Mn-Fe-O were obtained on extended areas of the substrate, see Fig.1. The chemical composition gradients of metallic constituents across the perovskite libraries are identified by EDX and the respective crystallographic phases by XRD. Since the La-content within the perovskite structure always stays close to its stoichiometry of 50 at. % with respect to the metal fractions, La-perovskite regions within the films can be identified by filtering the chemical composition data from the materials libraries for a certain La-content of interest. From empiric experience as well as from the known possible non-stoichiometry of La-based perovskites [16–20], this filter was set to comprise the La-contents of 45 at. % ≤ La ≤ 53 at. %. In **Figure 1**, photos of the sputter deposited La-Co-Mn-O and La-Co-Mn-Fe-O libraries are shown as well as the identified single-phase perovskite regions underneath each photo as a schematic representation of measurement areas. Gray squares symbolize measurement areas where neither the chemical composition, nor the analyzed phases give hint to single-phase perovskite structures. Orange squares display measurement areas where the chemical composition fits into the range of the chemical La-filter and therefore fulfills a precondition for single-phase perovskite. Squares with a dark blue dot are measurement areas where single-phase perovskite was confirmed by XRD phase analysis. For the subsequent analysis of perovskite properties and the correlation of resistivity and bandgap with the respective perovskite composition, only measurement areas are taken into account, that were identified as a single-phase by the La threshold values as well as by XRD (orange squares with blue dots). Especially measurement areas which show La content values below or above the aforementioned thresholds, i.e., below 45 at. % or above 53 at. %, but were identified as perovskite with XRD, are neglected, since these areas might contain some X-ray amorphous secondary phases, which cannot be identified by XRD.

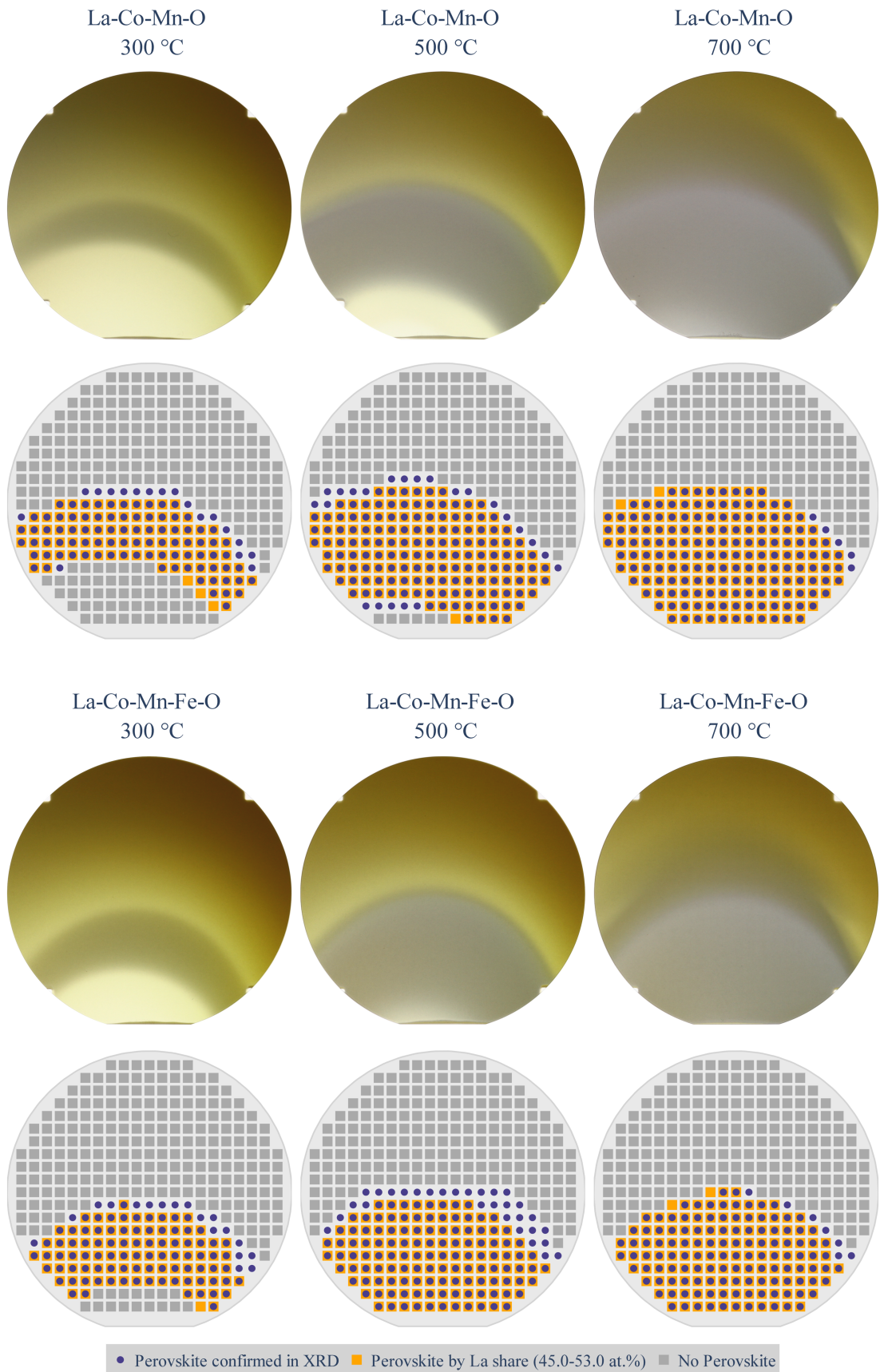

*Figure 1: Images of La-Co-Mn-O and La-Co-Mn-Fe-O libraries deposited at 300 °C, 500 °C, and 700 °C substrate temperature on double-sided polished sapphire wafers (c-plane orientation, 100 mm diameter). Below each image, schemes of the materials libraries are depicted, showing measurement areas where perovskite structure could either be identified by XRD (blue dots), by La content (orange squares) or could not be confirmed (gray squares).*

When the identified single-phase regions of the measurement areas are compared with photos of the materials libraries, one can see, that they perfectly match with the arcuate shaped grayish region of the materials libraries. Gray is the characteristic color of La-based Mn-rich perovskites like $LaMnO_3$[21]. That makes it easy to already identify the approximate regions of single-phase perovskite from the visual data from the photo documentation. Moreover, the spatial expansion of the single-phase perovskite region of the libraries increases with increasing substrate temperature for both materials systems.

Within the parts of the materials libraries with single-phase perovskite, the metal composition changes as shown in **Table 1**. For the La-Co-Mn-O system, the La-content fluctuates within the single-phase perovskite region about 7 at. %, the Co-content about 12 at. % and the Mn-content between 13-16 at. %. For the La-Co-Mn-Fe-O system, the La-content varies within the single-phase perovskite region by 4-7 at. %, the Co-content between 5-6 at. %, the Mn-content between 15-16 at. % and the Fe-content about 13 at. %.

*Table 1: Chemical composition range of single-phase perovskite regions in the La-Co-Mn-O and La-Co-Mn-Fe-O systems for different substrate temperatures during deposition, ranging from 300-700 °C.*

| System | Temperature | La-range | Co-range | Mn-range | Fe-range |
|---|---|---|---|---|---|
| **La-Co-Mn-O** | 300 °C | 45 – 52 at. % | 15 – 27 at. % | 24 – 40 at. % | --- |
| **La-Co-Mn-O** | 500 °C | 45 – 52 at. % | 11 – 23 at. % | 27 – 42 at. % | --- |
| **La-Co-Mn-O** | 700 °C | 45 – 52 at. % | 10 – 22 at. % | 29 – 42 at. % | --- |
| **La-Co-Mn-Fe-O** | 300 °C | 45 – 52 at. % | 13 – 18 at. % | 18 – 34 at. % | 7 – 20 at. % |
| **La-Co-Mn-Fe-O** | 500 °C | 45 – 50 at. % | 9 – 14 at. % | 22 – 38 at. % | 5 – 18 at. % |
| **La-Co-Mn-Fe-O** | 700 °C | 46 – 50 at. % | 8 – 14 at. % | 23 – 38 at. % | 5 – 18 at. % |

The filtered data reveals the large compositional space of single-phase perovskite, which is covered by the combinatorial sputter deposition approach

## 2.2 Surface microstructure and crystallinity

Microstructure as well as crystallinity and crystallite size of the single-phase perovskite region of both systems are changing with increasing deposition temperature. Crystallinity is discussed on the basis of the peak intensity and peak width of the XRD pattern of the film: narrow, intense peaks represent a higher crystallinity, than broad peaks with low intensity. Changes of the XRD-signal and the microstructure of the perovskite phase from the La-Co-Mn-O system with an approximate composition of $La(Co_{0.3}Mn_{0.7})O_{3\pm\sigma}$ are shown in **Figure 2** and from the La-Co-Mn-Fe-O system with an approximate composition of $La(Co_{0.15}Mn_{0.7}Fe_{0.15})O_{3\pm\sigma}$ are shown in **Figure** 3, respectively.

At 300 °C, the microstructure of the La(Co$_{0.3}$Mn$_{0.7}$)O$_{3\pm\delta}$ has a dune-like appearance and no distinct crystallites can be observed in SEM or AFM images. XRD measurement data only allows the identification of three broad peaks with low intensity, which belong to single-phase La(Co$_{0.3}$Mn$_{0.7}$)O$_{3\pm\delta}$ perovskite. At 500 °C, the microstructure of the perovskite still has a dune-like appearance, however crystallites were observed in the SEM and AFM images. The corresponding XRD pattern shows additional peaks, which also belong to single-phase La(Co$_{0.3}$Mn$_{0.7}$)O$_{3\pm\delta}$ and the peaks are narrower and have increased intensity. At 700 °C the microstructure of the perovskite consists of larger crystallites with well-defined borders. All peaks are more pronounced compared to the perovskite films deposited at 500 °C and gain in intensity. Therefore, one can reasonably assume an increase in crystallinity and crystallite size with increasing substrate temperature.

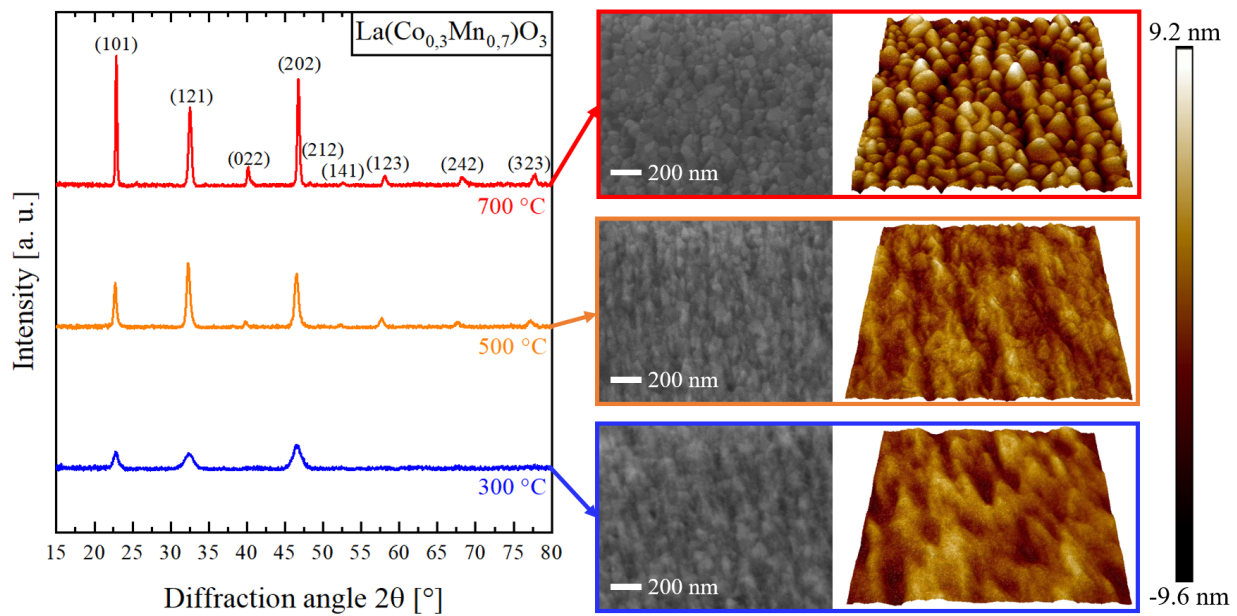

*Figure 2: Temperature-dependent evolution of the diffraction signal and the surface microstructure of perovskite with the approximate composition of La(Co$_{0.3}$Mn$_{0.7}$)O$_3$. Films were deposited on sapphire (c-plane) at temperatures ranging from 300-700 °C. SEM images were taken with 80k magnification and AFM images display a scanned area of 1µm×1µm. XRD peak indexing was done using ICSD-5960 and ICSD-5969.*

The La(Co$_{0.15}$Mn$_{0.7}$Fe$_{0.15}$)O$_{3\pm\delta}$ perovskite shows the same trends and a comparable evolution of crystallinity, crystallite size and microstructure as the La(Co$_{0.3}$Mn$_{0.7}$)O$_{3\pm\delta}$ perovskite. At 300 °C, it appears also dune-like and no large crystallites can be identified from SEM or AFM images. Again, on the XRD diffractogram three broad peaks with low intensity can be indexed, which belong to single-phase La(Co$_{0.15}$Mn$_{0.7}$Fe$_{0.15}$)O$_{3\pm\delta}$ perovskite. At 500 °C, the microstructure of a perovskite with equal composition still has a dune-like appearance, yet crystallites can be recognized from the SEM and AFM images. The corresponding XRD pattern shows additional peaks, which can be assigned to single-phase La(Co$_{0.15}$Mn$_{0.7}$Fe$_{0.15}$)O$_{3\pm\delta}$ and the peaks become more distinct and are generally increasing in intensity. At 700 °C the microstructure of the perovskite consists of larger crystallites with well-defined borders, that can easily be recognized from SEM as well as AFM images and are comparable to the ones observed for the La(Co$_{0.3}$Mn$_{0.7}$)O$_{3\pm\delta}$ perovskite.

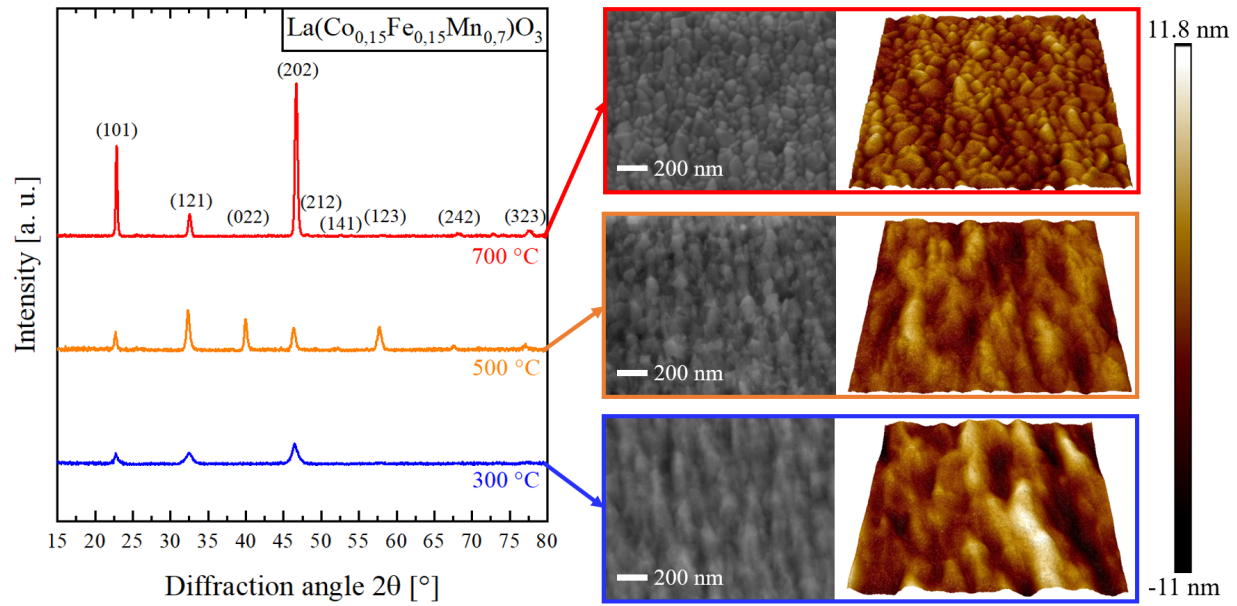

Figure 3: Temperature-dependent evolution of the diffraction signal and the surface microstructure of perovskite with the approximate composition of La(Co$_{0.15}$Mn$_{0.7}$Fe$_{0.15}$)O$_3$. Films were deposited on sapphire (c-plane) at temperatures ranging from 300-700 °C. SEM images were taken with 80k magnification and AFM images display a scanned area of 1µm×1µm. Peak indexing was done using ICSD-29036, ICSD-37296/7, ICSD-5960 and ICSD-5969.

For both systems, crystallinity and crystallite size increase with increasing substrate temperature resulting in a coarsening of the film microstructure.

From the perovskite region La(Co$_{0.15}$Mn$_{0.7}$Fe$_{0.15}$)O$_{3\pm\delta}$ deposited at 300 °C, a TEM lamella was prepared to confirm the single-phase nature of the perovskite regions of the thin films. A sample of this ML was chosen, since it is most prone to segregation and formation of additional phases, which might be X-ray amorphous, due to the higher chemical complexity and the rather low deposition temperature. The results are shown in **Figure 4**. Figure 4 a) shows a TEM image, where the marked area in red was analyzed further with EDX and high resolution (HR) images for FFT phase analysis. Figure 4 b) shows the results of an EDX mapping normalized to La, Co, Mn and Fe. The color gradient on the EDX maps is linked to a thickness gradient (thinner top to thicker bottom), which is already visible on Figure 4 a). No segregation was detected. Figure 4 c) shows the HR image and the corresponding fast Fourier transform (FFT) pattern (inset) of the area marked in the yellow square. The interplanar distances were measured and matched the reference for a La(Co$_{0.5}$Mn$_{0.5}$)O$_3$ perovskite (ICSD 5981) well, as shown in the table in Figure 4. No evidence for other crystallographic structures or amorphous regions were found. Combined with the EDX results in Figure 4 b), this confirms the lamella to be single-phase perovskite and strongly suggests that all other perovskite regions—identified by XRD and La composition limits—are indeed single-phase perovskite.

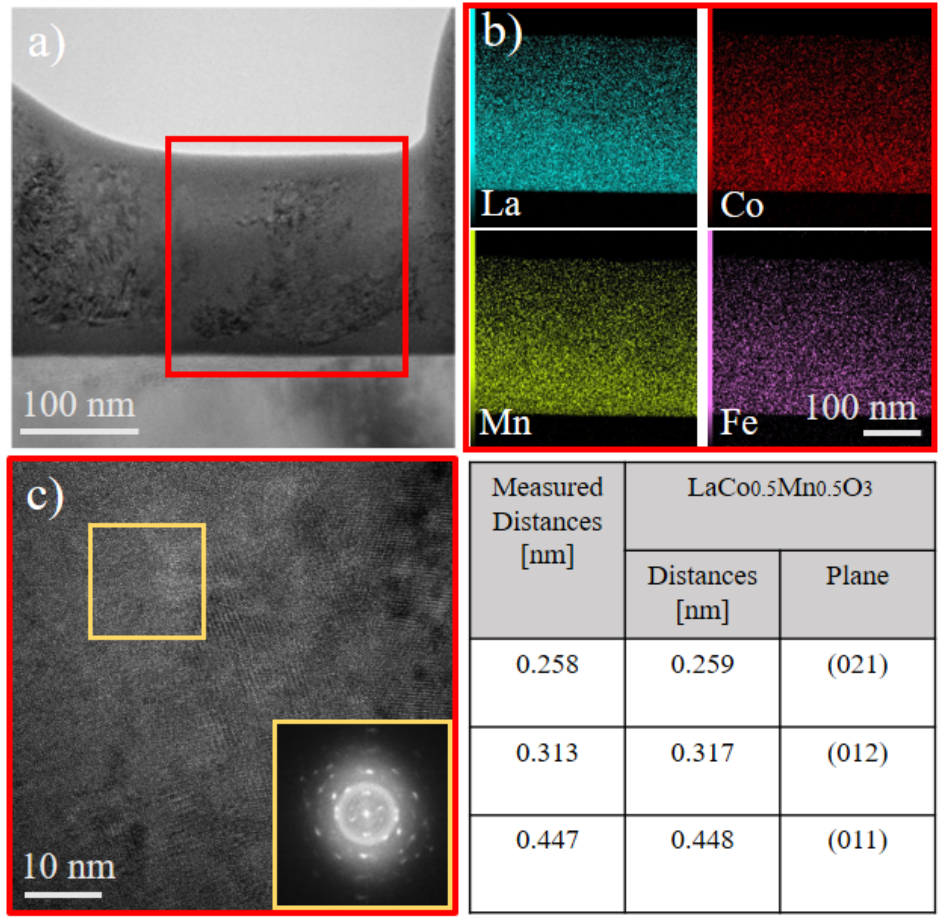

*Figure 4: Results of TEM measurements on a cross-sectional sample prepared from the perovskite region of the La-Co-Mn-Fe-O system deposited at 300 °C. a) TEM image, the analysis region for the following is marked in red, b) EDX composition maps normalized to La, Co, Mn, Fe. The color gradient is related to the lamella thickness, which increases from top to bottom. C) HR TEM image and corresponding FFT (inset), adjacent a table with the measured interplanar distances in nm compared to a perovskite reference (ICSD 5981).*

## 2.3 High-throughput characterization of optical bandgap values

The optical bandgap of La-based perovskites is dependent on their chemical composition. Literature values for pure $LaCoO_3$ are in the range of 2.00 eV to 2.41 eV[22–25], for pure $LaMnO_3$ in the range of 1.6 eV up to 3 eV[26–29] and for pure $LaFeO_3$ in the range of 2.15eV to 2.5 eV[30–32]. The observed variation in bandgap energy within one material, might be caused by different factors. Different synthesis methods, residues from the synthesis, amorphous phases, crystallinity as well as inaccurate performed Tauc-plots can lead to deviations in determined bandgap energies. Due to the high purity films prepared by sputtering, residues from the synthesis are avoided and by using an automated Tauc-Plot, as described in the experimental section of this article, a possible inaccuracy due to subjective errors is minimized.

For the investigated perovskite libraries, a dependency of bandgap energy from the perovskite composition was observed, as shown in **Figure 5** a) and c) and **Table 2**. With increasing Mn-content in the perovskite, bandgap energy was increasing continuously for both perovskite systems. Increasing amounts of Co and Fe appear to reduce the bandgap energy. The determined values are in agreement with literature [22,33] Table 2 summarizes the perovskite compositions with the minimum and maximum bandgap energies measured within both materials systems, deposited at 300 °C, 500 °C and 700 °C, respectively.

*Table 2: Overview over the perovskite compositions with the minimum and maximum bandgap energies from the La-Co-Mn-O and the La-Co-Mn-Fe-O systems deposited at 300 °C, 500 °C and 700 °C.*

| Temperature [°C] | Material | $E_g$ [eV] |
|---|---|---|
| 300 | $La_{0.48}(Co_{0.26}Mn_{0.26})O_{3\pm\delta}$ | 2.54 |
| | $La_{0.48}(Co_{0.17}Mn_{0.35})O_{3\pm\delta}$ | 2.72 |
| 500 | $La_{0.49}(Co_{0.20}Mn_{0.31})O_{3\pm\delta}$ | 2.54 |
| | $La_{0.48}(Co_{0.12}Mn_{0.40})O_{3\pm\delta}$ | 2.78 |
| 700 | $La_{0.47}(Co_{0.22}Mn_{0.31})O_{3\pm\delta}$ | 2.62 |
| | $La_{0.49}(Co_{0.11}Mn_{0.40})O_{3\pm\delta}$ | 2.86 |
| 300 | $La_{0.47}(Co_{0.16}Mn_{0.19}Fe_{0.18})O_{3\pm\delta}$ | 2.63 |
| | $La_{0.52}(Co_{0.15}Mn_{0.19}Fe_{0.14})O_{3\pm\delta}$ | 2.75 |
| 500 | $La_{0.48}(Co_{0.13}Mn_{0.22}Fe_{0.17})O_{3\pm\delta}$ | 2.66 |
| | $La_{0.49}(Co_{0.11}Mn_{0.30}Fe_{0.10})O_{3\pm\delta}$ | 2.78 |
| 700 | $La_{0.48}(Co_{0.10}Mn_{0.27}Fe_{0.15})O_{3\pm\delta}$ | 2.67 |
| | $La_{0.49}(Co_{0.09}Mn_{0.35}Fe_{0.07})O_{3\pm\delta}$ | 2.84 |

Regarding a possible influence of the substrate temperature, it was not possible to identify a clear trend or difference, since the perovskite libraries do not cover the exact same compositional space. Moreover, the observed deviations at the different temperatures are negligibly small.

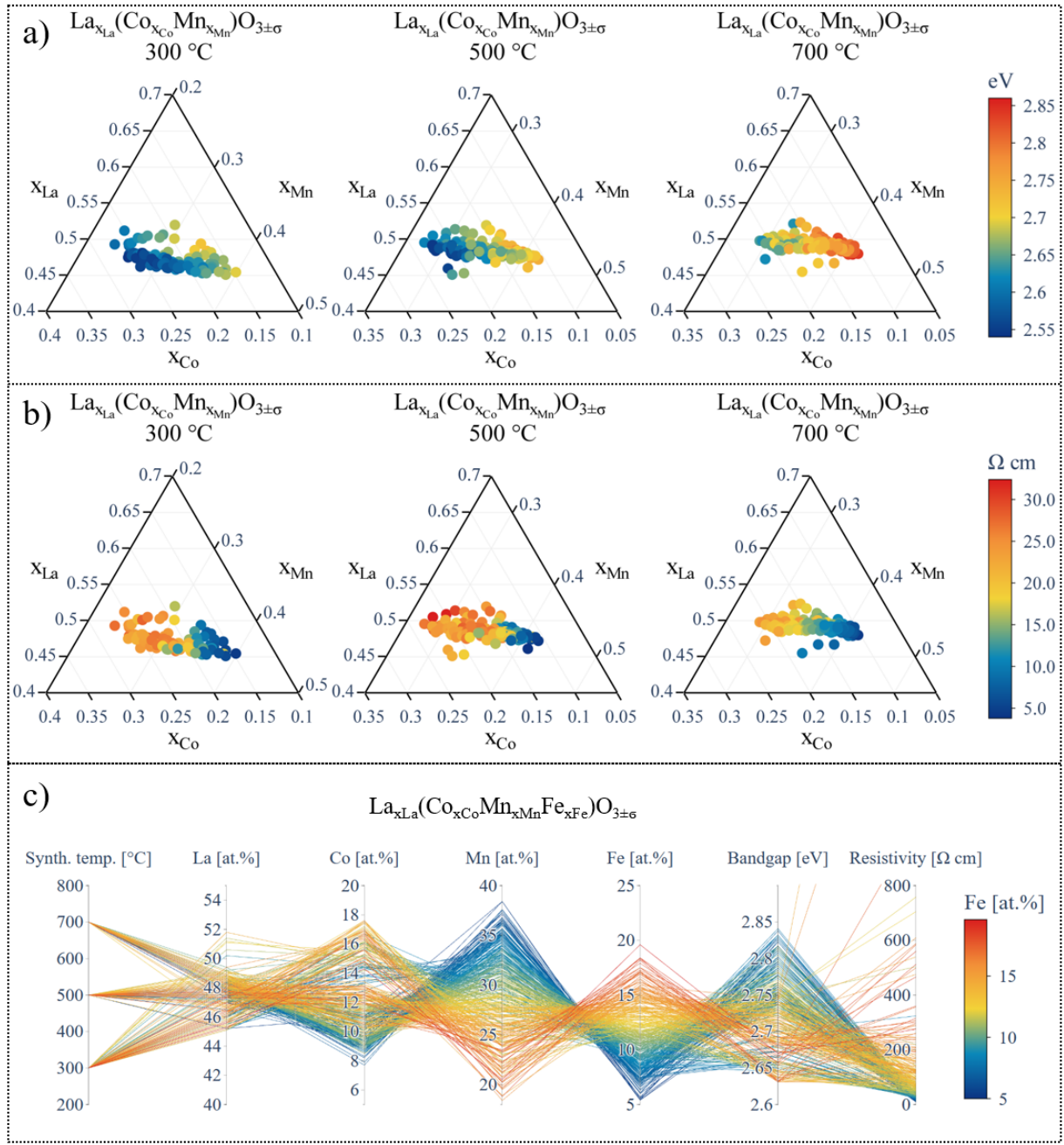

*Figure 5: Results of high-throughput characterization of the optical bandgap values and electrical resistivity of $La_{x_{La}}(Co_{x_{Co}}Mn_{x_{Mn}})O_{3\pm\sigma}$ and $La_{x_{La}}(Co_{x_{Co}}Mn_{x_{Mn}}Fe_{x_{Fe}})O_{3\pm\sigma}$ for the respective measurement areas of all samples deposited at 300 °C, 500 °C and 700 °C substrate temperature: composition- and temperature-dependent (a) optical bandgap energies, (b) electrical resistivity values of $La_{x_{La}}(Co_{x_{Co}}Mn_{x_{Mn}})O_{3\pm\sigma}$ and (c) optical bandgap values as well as electrical resistivity of $La_{x_{La}}(Co_{x_{Co}}Mn_{x_{Mn}}Fe_{x_{Fe}})O_{3\pm\sigma}$ in parallel coordinates. This type of plot is used, since the representation of four-component systems in a ternary diagram is not feasible. One dataset, consisting of synthesis temperature, chemical composition, bandgap and resistivity, is thereby one line crossing all dimensional axes. The colormap is varying according to the Fe content in at.%.*

## 2.4 High-throughput characterization of electrical resistivity

Screening of the electrical resistivity of the libraries reveals a dependency of resistivity from the perovskite composition in both materials systems as shown in Figure 5 b) and c) and **Table 3**. For $La(Co_{1-x}Mn_x)O_3$ perovskites this compositional dependency of the resistivity is documented in literature [34–36]: The perovskite with the approximate composition of $LaCo_{0.5}Mn_{0.5}O_3$ is known to have the highest resistivity, with documented values between 40 Ωcm to 60 Ωcm. With increasing as well as decreasing Co-content in the perovskite, resistivity decreases, in agreement with the trends and resistivity values observed in this work. If the $La(Co_{1-x}Mn_x)O_3$ perovskite is doped or substituted with larger amounts of Fe, the electrical resistivity of the perovskite increases significantly[37,38]. The same effect was observed for the Fe-substituted perovskite in this study. With increasing Fe-content, electrical resistivity was increasing significantly, as shown in Figure 5 c).

*Table 3: Overview over the perovskite compositions with the minimum and maximum electrical resistivity from the La-Co-Mn-O and the La-Co-Mn-Fe-O systems deposited at 300 °C, 500 °C and 700 °C.*

| Temperature [°C] | Material | $\rho$ [Ω cm] |
|---|---|---|
| 300 | $La_{0.45}(Co_{0.15}Mn_{0.40})O_{3\pm\delta}$ | 4 |
| 300 | $La_{0.48}(Co_{0.24}Mn_{0.28})O_{3\pm\delta}$ | 28 |
| 500 | $La_{0.46}(Co_{0.13}Mn_{0.41})O_{3\pm\delta}$ | 4 |
| 500 | $La_{0.50}(Co_{0.22}Mn_{0.28})O_{3\pm\delta}$ | 32 |
| 700 | $La_{0.48}(Co_{0.10}Mn_{0.42})O_{3\pm\delta}$ | 4 |
| 700 | $La_{0.49}(Co_{0.18}Mn_{0.33})O_{3\pm\delta}$ | 25 |
| 300 | $La_{0.45}(Co_{0.13}Mn_{0.34}Fe_{0.08})O_{3\pm\delta}$ | 22 |
| 300 | $La_{0.52}(Co_{0.15}Mn_{0.19}Fe_{0.14})O_{3\pm\delta}$ | 3027 |
| 500 | $La_{0.48}(Co_{0.10}Mn_{0.36}Fe_{0.06})O_{3\pm\delta}$ | 9 |
| 500 | $La_{0.49}(Co_{0.11}Mn_{0.30}Fe_{0.10})O_{3\pm\delta}$ | 1022 |
| 700 | $La_{0.48}(Co_{0.08}Mn_{0.38}Fe_{0.06})O_{3\pm\delta}$ | 11 |
| 700 | $La_{0.49}(Co_{0.09}Mn_{0.35}Fe_{0.07})O_{3\pm\delta}$ | 307 |

For the La-Co-Mn-O system no influence of the substrate temperature on the electrical resistivity was observed. As for the bandgap energies, it is not possible to identify a clear trend or difference, since the perovskite libraries do not cover the exact same compositional space. In case of the La-Co-Mn-Fe system, substrate temperature appears to have an impact on the resistivity. With increasing substrate temperature during deposition, resistivity decreases. Although the perovskite libraries do not cover the exact same compositional space, the difference between the values is so high (about factor 10), that a temperature dependence can be concluded. Possible reasons for the difference might be changes in crystallinity, crystallite size, oxygen fraction or defects. However, the exact cause could not be identified and is beyond the scope of this work.

# 3 Conclusion

The single-phase formation behavior for reactive co-sputter deposition of La-based materials systems was used for combinatorial perovskite screening on example of the systems La-Co-Mn-O and La-Co-Mn-Fe-O: at deposition temperatures ≥ 300 °C single-phase perovskite, which continuously varies its composition, forms on extended regions of the substrate. With increasing substrate temperature, the area of the single-phase perovskite region increases. Moreover, the crystallinity of the perovskite increases with temperature for both systems. Screening of the libraries reveals a compositional dependence of bandgap energy as well as resistivity for both perovskite systems. In the $La_{x_{La}}(Co_{x_{Co}}Mn_{x_{Mn}})O_{3\pm\delta}$ perovskite, bandgap energy increases with increasing Mn-content, whereas resistivity decreases with increasing Mn-content. Addition of Fe in the $La_{x_{La}}(Co_{x_{Co}}Mn_{x_{Mn}})O_{3\pm\delta}$ perovskite leads to an increase of resistivity with increasing Fe-content, while bandgap energy decreases with increasing Fe-content. All determined values and trends are in agreement with values known in literature. This emphasizes the benefit of the phase formation phenomenon as foundation for the application in combinatorial perovskite screening and offers a new efficient, time and cost saving scientific approach for the field of perovskite research.

# 4 Experimental methods

## 4.1 Synthesis of La-Co-O thin film libraries

La-Co-Mn-O and La-Co-Mn-Fe-O libraries were fabricated by reactive ($O_2$/Ar atmosphere) magnetron co-sputtering in a commercial four-cathode sputter system (AJA International, ATC 2200 with four confocally aligned sputter sources). Within this system, the cathodes are arranged circular around the substrate with an angle of 90° of the neighboring cathodes to each other, as shown in **Figure 6**. La was sputtered from a $La_2O_3$ compound target (4-inch diameter, 99.99 % purity, Evochem Advanced Materials GmbH), Co from an elemental Co-Target (4-inch diameter, 99.99 %, Sindlhauser Materials GmbH), Mn from an elemental Mn-target (4-inch diameter, 99.95 %, Sindlhauser Materials GmbH) and Fe from an elemental Fe-target (4-inch diameter, 99.99 %, Sindlhauser Materials GmbH).

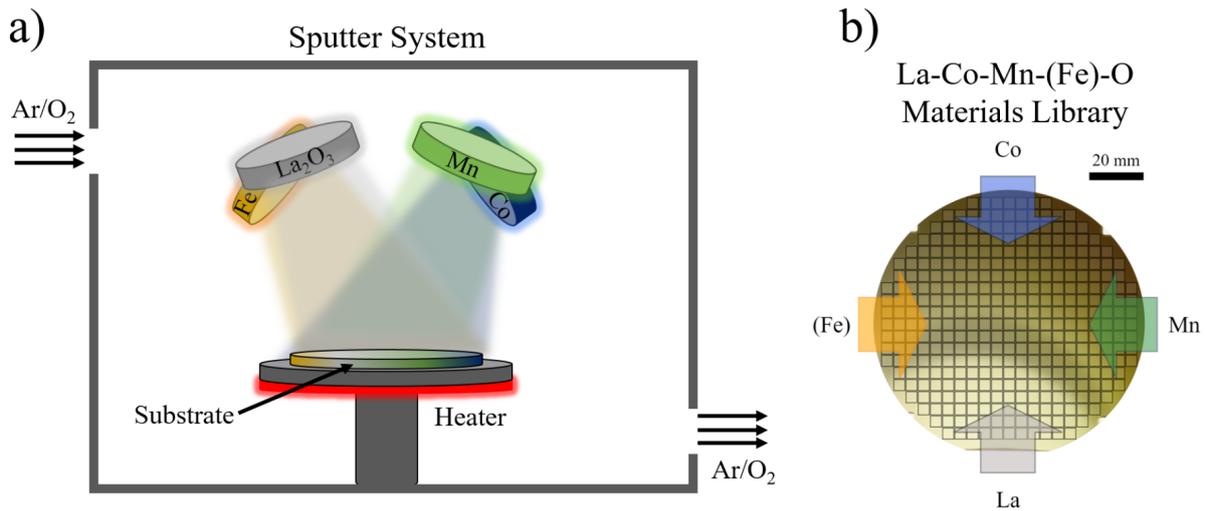

*Figure 6: a) Schematic illustration of the reactive magnetron co-deposition process to fabricate materials libraries (MLs) of ternary or quaternary oxide thin film systems. b) A schematic of the ML deposited on a 4-inch diameter wafer is shown and the sputter target positions are indicated.*

For all depositions, the cathode configuration was fixed: The $La_2O_3$- and the Co-target were positioned 180° to each other as well as the Fe- and Mn-target (see Figure 6). For the La-Co-Mn-O libraries, $La_2O_3$ was sputtered with a RF power of 200 W, Co was sputtered with a DC power of 70 W and Mn sputtered with a pulsed DC power of 130 W (10 kHz, 5 µs). For the La-Co-Mn-Fe-O libraries the same powers were used and additionally Fe was co-sputtered with a DC power of 130 W. All used process gasses had 6N purity (99.9999%). The Ar line has an additional getter: SAES pure gas type MC1-903T purifier that typically removes $O_2$, $H_2O$, CO, $CO_2$ and organic/NMHC compounds <100 ppt and a 0.003 µm particle filter. There was no getter purifier on the $O_2$ gas line. The base pressure at all depositions was in the low range of $10^{-5}$ Pa and depositions were carried out at a fixed process pressure of 0.4 Pa with a constant $O_2$/Ar flow ratio of 40 sccm/80 sccm. Film thickness was controlled by deposition time. As substrates 100 mm diameter sapphire wafers (c-plane, SITUS Technicals GmbH) were used. For hot depositions, substrate temperatures of 300 °C, 500 °C and 700 °C were used. After finishing each deposition, all cathodes were powered off, gas flux was switched off and the heater was shut down. So, all hot deposited materials libraries cooled down under vacuum and did not experience any further intentional heat- or oxidation treatment.

## 4.2 High-throughput characterization

High-throughput means the automated, consecutive characterization of the ML with suitable automated methods, following a predefined identical measurement scheme across a library. This way, all methods are applied to the same measurement areas (MAs), which makes the different results comparable to each other: Every 100 mm diameter library is divided into 342 contiguous measurement areas with an identical size of 4.5 mm × 4.5 mm. Every measurement is started in the center of the first MA and then continued in a fixed sequence to the center of the respective next MA with a step size of 4.5 mm until all 342 MAs are measured. The actual size of the measured areas depends on the used measurement method.

Film thickness was measured using a tactile profilometer (Ambios XP2) on steps of partially removed film. These steps were created using photolithographic techniques or by partially covering the substrate during film deposition. On selected samples, films thickness was additionally checked by SEM cross sections.

The chemical composition, with respect to the metals and omitting oxygen, was analyzed using automated energy-dispersive X-ray spectroscopy (EDX) in a scanning electron microscope (SEM, JEOL 5800 equipped with an Oxford INCA X-act detector). The SEM was operated at 20 kV with a magnification of 600, leading to a measured area of 400 μm × 600 μm. The composition of these MAs is averaged over the area and considered homogeneous within the MA. Before every measurement, the EDX system is calibrated on a Co-standard, which leads to a measurement accuracy of 1 at. %. Measurement data was normalized on La, Co and Mn for the La-Co-Mn-O system and La, Co, Mn and Fe for the La-Co-Mn-Fe-O system and is given in at. %. All compositions measured with EDX are described and discussed on the basis of the metallic gradient since the O-signal is biased by the oxide substrate and therefore cannot be quantified by EDX.

The crystallographic phase analysis was performed by X-ray diffraction (XRD). A Bruker D8 Discover with a Vantec-500 2D-detector in Bragg-Brentano geometry and an Incoatec High Brilliance Iμs Cu $K_\alpha$ X-ray source (0,15418 nm) was used. Three frames were taken stepwise at every MA with an increment of θ/2θ 10°/20°, starting at 12,5°/25° and finishing at 32.5°/65°. This way an angular 2θ range from approximately 10° to 85° was covered. For visualization, the angular 2θ range of the diffraction pattern was limited from 15° to 80°. For peak indication ICSD references were used. Since the oxygen content of the crystallographic phases was not determined, the stoichiometry within the phases was complemented with the term "±σ".

Electric resistance of the films at room temperature (RT) was determined by 4-point probe (4PP) resistance measurements, using an in-house built high-throughput measurement setup [39]. In this setup, a Keithley 2400 SourceMeter was used, which was operated in the automated mode, to dynamically change voltage and current depending on the thin film's resistance. The maximum resistance that can be measured with this setup is ≈ 200 MΩ. Electrical resistivity of the films was calculated from the electrical resistance values and the film thickness.

Optical bandgap energies were determined from high-throughput UV-vis optical transmission measurements using an in-house developed measurement stand [40,41]. In this setup, all MA are sequentially illuminated from the film side with a 90° angle of incidence from a Xe-lamp. The transmitted signal is detected and evaluated using a customized spectrometer (Ocean Optics Hr2000+) in a spectral range from 350 nm to 900 nm. From the measured transmission values, the absorbance was calculated which was further used to calculate the bandgap energy using the Tauc-plot method. To minimize errors due to manual determination, Tauc-plots were created automatically by an in-house written software following the paper from Suram et al. [42]

The surface morphology was examined using a SEM (JEOL JSM-7200F), operated at acceleration voltages in the range of 5 kV to 15 kV, depending on the thin film's conductivity, and with a working distance of about 5 mm. AFM images were recorded using a Bruker FastScan scanning probe microscope in PeakForce Tapping mode with ScanAsyst, using a ScanAsyst Air probe (spring constant approximately 0.4 N/m). On one sample additional TEM (JEOL JEM ARM200F) measurements, EDX and HR images for FFT, were carried out. These measurements were done to confirm that there is only the single-phase perovskite present in the thin films. The TEM lamella was prepared using a focused ion beam (FIB) system (Helios NanoLab G4 CX) with Ga ion beam.

## 4.3 Acknowledgements

This work was funded by the Deutsche Forschungsgemeinschaft (DFG, German Research Foundation) - Projektnummer 388390466-TRR 247, project C4.

Additionally, SFB TR 87 and SFB TR 103 are acknowledged for support.

ZGH at Ruhr-University Bochum is acknowledged for the use of SEM, FIB, TEM, XRD, AFM and the corresponding measurements. Dr. Aleksander Kostka is acknowledged for his support during TEM measurements.

The authors declare no competing interests.

# 5 Author Information

## 5.1 Corresponding Author

**Alfred Ludwig** - Chair for Materials Discovery and Interfaces, Institute for Materials, Ruhr University Bochum, 44780 Bochum, Germany; orcid.org/0000-0003-2802-6774; Email: alfred.ludwig@rub.de

## 5.2 Authors

**Tobias H. Piotrowiak** − Chair for Materials Discovery and Interfaces, Institute for Materials, Ruhr University Bochum, 44780 Bochum, Germany; https://orcid.org/0000-0001-9378-1851

**Rico Zehl** – Chair for Materials Discovery and Interfaces, Institute for Materials, Ruhr University Bochum, 44780 Bochum, Germany; https://orcid.org/0000-0003-2390-5913

**Ellen Suhr** – Chair for Materials Discovery and Interfaces, Institute for Materials, Ruhr University Bochum, 44780 Bochum, Germany; https://orcid.org/0000-0003-3461-8274

**Lars Banko** – Chair for Materials Discovery and Interfaces, Institute for Materials, Ruhr University Bochum, 44780 Bochum, Germany

**Benedikt Kohnen** – Chair for Materials Discovery and Interfaces, Institute for Materials, Ruhr University Bochum, 44780 Bochum, Germany